\documentclass[prd,preprint,nofootinbib]{revtex4}

\usepackage{graphicx}
\usepackage{amssymb}
\usepackage{amsmath}

\begin{document}

\renewcommand{\thefootnote}{\#\arabic{footnote}}
\newcommand{\rem}[1]{{\bf [#1]}}
\newcommand{\gsim}{ \mathop{}_ {\textstyle \sim}^{\textstyle >} }
\newcommand{\lsim}{ \mathop{}_ {\textstyle \sim}^{\textstyle <} }
\newcommand{\vev}[1]{ \left\langle {#1}  \right\rangle }
\newcommand{\bear}{\begin{array}}  
\newcommand {\eear}{\end{array}}
\newcommand{\bea}{\begin{eqnarray}}   
\newcommand{\eea}{\end{eqnarray}}
\newcommand{\beq}{\begin{equation}}   
\newcommand{\eeq}{\end{equation}}
\newcommand{\bef}{\begin{figure}}  
\newcommand {\eef}{\end{figure}}
\newcommand{\bec}{\begin{center}} 
\newcommand {\eec}{\end{center}}
\newcommand{\non}{\nonumber}  
\newcommand {\eqn}[1]{\beq {#1}\eeq}
\newcommand{\la}{\left\langle}  
\newcommand{\ra}{\right\rangle}
\newcommand{\ds}{\displaystyle}

\def\SEC#1{Sec.~\ref{#1}}
\def\FIG#1{Fig.~\ref{#1}}
\def\EQ#1{Eq.~(\ref{#1})}
\def\EQS#1{Eqs.~(\ref{#1})}
\def\REF#1{(\ref{#1})}
\def\lrf#1#2{ \left(\frac{#1}{#2}\right)}
\def\lrfp#1#2#3{ \left(\frac{#1}{#2} \right)^{#3}}
\def\GEV#1{10^{#1}{\rm\,GeV}}
\def\MEV#1{10^{#1}{\rm\,MeV}}
\def\KEV#1{10^{#1}{\rm\,keV}}

\def\lrf#1#2{ \left(\frac{#1}{#2}\right)}
\def\lrfp#1#2#3{ \left(\frac{#1}{#2} \right)^{#3}}

\begin{flushright}
IPMU 09-0052
\end{flushright}

\title{
Phenomenological Aspects of Ho\v{r}ava-Lifshitz Cosmology
}

\author{
Shinji Mukohyama$^{(a)}$,
Kazunori Nakayama$^{(b)}$,
Fuminobu Takahashi$^{(a)}$,
Shuichiro Yokoyama$^{(c)}$
}

\affiliation{%
$^a$ Institute for the Physics and Mathematics of the Universe, 
University of Tokyo, Chiba 277-8568, Japan\\
$^b$ Institute for Cosmic Ray Research,
University of Tokyo, Kashiwa 277-8582, Japan,\\
$^c$ Department of Physics and Astrophysics, Nagoya University, Aichi 464-8502, Japan
}

\date{\today}

\begin{abstract}
We show that, assuming the dispersion relation proposed recently
by Ho\v{r}ava in the context of quantum gravity, radiation energy density
exhibits a peculiar dependence on the scale factor; the radiation energy
density decreases proportional to $a^{-6}$. This simple scaling can have
an impact on cosmology. As an example, we show that the resultant baryon
asymmetry as well as the stochastic gravity waves can be enhanced. We also discuss
current observational constraint on the dispersion relation.
\end{abstract}


\maketitle

\section{Introduction}
\label{sec:1}

Recently a new class of quantum gravity was proposed by
Ho\v{r}ava~\cite{Horava:2009uw},  motivated by the solid state
physics. This theory is fundamentally non-relativistic and, in the
ultraviolet (UV), exhibits the Lifshitz scale invariance 
%
\begin{equation}
 t \to b^z t, \quad \vec{x}\to b \vec{x},
  \label{eqn:scaling}
\end{equation}
with dynamical critical exponent $z=3$. It is this anisotropic rescaling
that makes Ho\v{r}ava's theory power-counting
renormalizable. The relativistic scaling with $z=1$ is restored in the 
infrared (IR) due to deformation by relevant operators, and the Lorentz
symmetry emerges as an accidental symmetry.

The $z=3$ scaling implies that, in the UV, a physical degree of freedom
should have a dispersion relation of the form 
%
\begin{equation}
 \omega^2 \;\simeq\; \frac{k^6}{M^4},
  \label{eqn:dispersion-relation-UV}
\end{equation}
where $M$ is a characteristic mass scale. Since the Planck
mass is an emergent quantity in Ho\v{r}ava's theory, $M$ does not have
to be the same order of magnitude as the Planck mass. Moreover, $M$ can
differ for different physical degrees of freedom: graviton, photon,
quarks, leptons, scalar fields, etc. The fundamental symmetry of
Ho\v{r}ava's theory is invariance under foliation-preserving
diffeomorphism: 
%
\begin{equation}
 x^i \to \tilde{x}^i(x^j,t), \quad t\to \tilde{t}(t).
\end{equation}
Any value of $M$ is consistent with this symmetry and, thus, there is no
symmetry reason to expect any fundamental relations among $M$'s for
different species unless additional assumptions are made.

On the other hand, among relevant deformations to the dispersion
relation (\ref{eqn:dispersion-relation-UV}), the coefficient of the
$k^2$ term should flow to the squared speed of light at the IR fixed
point. (See \EQ{dis-rel}.) This applies to all species, since relativity would not
emerge in the IR, otherwise.

Various aspects of Ho\v{r}ava-Lifshitz gravity were already explored in
the literature. For example, cosmology based on Ho\v{r}ava-Lifshitz
gravity was investigated
in~\cite{Takahashi:2009wc,Calcagni:2009ar,Kiritsis:2009sh,Mukohyama:2009gg,Brandenberger:2009yt,Nikolic:2009jg,Piao:2009ax,Gao:2009bx,Chen:2009ka}. 
Some solutions in Ho\v{r}ava-Lifshitz gravity were presented
in~\cite{Lu:2009em,Nastase:2009nk,Cai:2009pe,Colgain:2009fe}. 
Possible extensions of the theory were
proposed~\cite{Kluson:2009sm,Izawa:2009ne,Cai:2009ar,Sotiriou:2009gy}.

The dispersion relation (\ref{eqn:dispersion-relation-UV}) leads to
interesting cosmological consequences, such as generation of
scale-invariant cosmological perturbations~\cite{Mukohyama:2009gg} and
time-delays in Gamma-ray bursts~\cite{Chen:2009ka}. The purpose of this
paper is to explore yet another cosmological consequence of the $z=3$
dispersion relation. We shall point out that the radiation energy
density in the UV epoch ($T\gg M$) is proportional to $a^{-6}$ and,
thus, decays faster than in the IR epoch ($T\ll M$) or in relativistic
theories. This leads to intriguing cosmological consequences such as
enhancement of baryon asymmetry and stochastic gravity waves. We shall
also discuss current observational constrains on the dispersion relation.

This paper is organized as follows. 
In Sec.~\ref{sec:2} cosmological impacts of dispersion relation (\ref{eqn:dispersion-relation-UV})
are explained with some explicit examples.
In Sec.~\ref{sec:3} observational constraints on the dispersion relation, 
in particular, the value of $M$ is discussed.
Sec.~\ref{sec:4} is devoted to summary.

\section{Cosmological impact of dispersion relation}
\label{sec:2}
We first derive the peculiar dependence of radiation energy density
on the scale factor of the Universe. Suppose that the constituent particles
satisfy the dispersion relation,
\beq
\omega^2 \;\simeq\; \frac{k^6}{M^4} + \kappa \frac{k^4}{M^2} + k^2,
\label{dis-rel}
\eeq
where we set the speed of light $c$ to be unity, for simplicity.
If the typical momentum of the particle ($\sim$ temperature) is much lower than $M$,
we recover the ordinary dispersion relation, $\omega^2 \simeq k^2$. On the other hand,
if the momentum is much larger than $M$, the dispersion relation can be approximated to be
\beq
\omega^2 \;\simeq\; \frac{k^6}{M^4}.
\label{dis-rel2}
\eeq
Assuming the relation \REF{dis-rel2}, the energy density of the radiation, $\rho_r$, decreases as
\beq
\rho_r \sim \omega n \propto a^{-6},
\label{a-6}
\eeq
where $\omega$ is a typical energy of the particles, $n$ the number density,
and $a$ represents the scale factor.
This corresponds to the fluid with the equation of state $p_r=\rho_r$
where $p_r$ is the pressure of the radiation.
Here we have used  a fact that the physical momentum
$k$ redshifts as $k \propto a^{-1}$ due to the cosmic expansion, and the number
density decreases as $n \propto a^{-3}$. The peculiar dependence \REF{a-6}
should be contrasted to the ordinary scaling, $\rho_r \propto a^{-4}$.
Since the energy density of the radiation decreases more quickly than the ordinary
case, the estimate on cosmological abundance of e.g. the baryon number, the stochastic gravity waves, etc.
can be modified. Let us see this below.

\subsection{Baryon asymmetry}

As an example we take up non-thermal baryon production from an inflaton decay.\footnote{
We call a field as inflaton, which dominates the energy density of the
Universe and decays into radiation. It does not affect our discussion whether or not
the field actually induces the inflationary expansion.}
Suppose that, after inflation ends, the inflaton $\phi$ decays into radiation, which is heated up to 
a temperature higher than $M$. At the same time, the baryon asymmetry is assumed to be
generated from the inflaton decay. This is indeed the case of non-thermal leptogenesis
\cite{Fukugita:1986hr,Murayama:1993em,Asaka:1999yd}. 
As the Universe expands, the temperature will decrease.
We assume that the ordinary dispersion, $\omega = k$,
is restored at a temperature below $M$.  Then, the final  baryon-to-entropy ratio can be estimated as
\bea
\left.\frac{n_B}{s}\right|_{T=M} &=& \left.\frac{\sqrt{\rho_r}}{s}\right|_{T=M}\left.\frac{n_B}{\sqrt{\rho_r}}\right|_{T=M}
= \left.\frac{\sqrt{\rho_r}}{s}\right|_{T=M}\left.\left(\frac{\rho_\phi}{\sqrt{\rho_r}}\frac{n_B}{\rho_\phi}\right)\right|_{Reheating},
\eea
where 
$\rho_\phi$ denotes the inflaton energy density.\footnote{Note that the entropy density is proportional to $T^3$ for $T < M$, 
but this could be modified for $T > M$. The calculation here does not depend on the definition of the entropy for $T>M$.}
 The final baryon asymmetry will be
\bea
\left.\frac{n_B}{s}\right|_{T=M} &\sim&\frac{\varepsilon}{M} \frac{\Gamma_\phi M_P}{m_\phi},
\eea
where  $M_P \simeq 2.4 \times 10^{18}$\,GeV is the reduced Planck mass,
$\Gamma_\phi$ denotes the decay rate of the inflaton, and
$\varepsilon$ is the baryon asymmetry generated from the decay
of one inflaton quanta. This result should be compared with the ordinary result,
\beq
\frac{n_B}{s} \;\sim\; \varepsilon\frac{\sqrt{\Gamma_\phi M_P}}{m_\phi}.
\eeq
Thus, we conclude that the baryon asymmetry is enhanced by 
\beq
\Delta \sim \frac{\sqrt{\Gamma_\phi M_P}}{M},
\eeq
compared to the case of the ordinary dispersion relation. Such an enhancement may make
an inefficient baryogenesis scenario viable,  which would result in too small baryon asymmetry, otherwise.

\subsection{Gravitational waves}

Following the similar arguments, it is also possible to enhance abundance of the relic gravity waves.
Although the amplitude of the gravitational waves remain constant when the corresponding mode
lies outside the horizon, once it enters the horizon its amplitude decreases 
inversely proportional to the scale factor.
Since the timing at which a mode enters the horizon depends on the equation of state of the Universe,
the thermal history of the Universe is imprinted in the gravitational wave spectrum
\cite{Seto:2003kc,Tashiro:2003qp,Boyle:2005se,Nakayama:2008ip,Kuroyanagi:2008ye}.
The present spectrum of the stochastic gravitational wave background in terms of the density parameter 
$\Omega_{\rm gw}(k)$ is given by
\beq
	\Omega_{\rm gw}(k) \equiv \frac{1}{\rho_{c0}}\frac{d\rho_{\rm gw}}{d\ln k}
	= \frac{1}{12H_0^2}k^2 {\Delta_h^{(\rm prim)}(k)}^2
	\left( \frac{a_{\rm in}(k)}{a_0} \right)^2,
\eeq
where $\rho_{c0}$ is the critical energy density at present, 
$d\rho_{\rm gw}/d\ln k$ represents the energy density of the gravitational waves per logarithmic
frequency, $H_0$ is the present Hubble parameter,
$a_{\rm in}(k)$ denotes the scale factor when the mode with comoving wave number $k$
enters the horizon, and $\Delta_h^{(\rm prim)}(k)$ is the 
dimensionless power spectrum of the primordial gravitational waves,
which is assumed to be scale invariant.
From this equation we can see that
the resulting spectrum scales as $\Omega_{\rm gw}(k) \propto k^{(2-4p)/(1-p)}$
if the Universe expands as $a\propto t^p$ when the mode $k$ enters the horizon.
Therefore we get
\beq
	\Omega_{\rm gw}(k) \propto \left \{
		\begin{array}{ll}
			k^{-2} &{\rm for}~~p=2/3~(w=0), \\
			k^{0} &{\rm for}~~p=1/2~(w=1/3), \\
			k^{1} &{\rm for}~~p=1/3~(w=1), 
		\end{array}
	\right. 
\eeq
where $w$ represents the equation of state of the Universe.
The last line corresponds to the case where the radiation in the UV regime fills the Universe.
Thus the mode with $k>k_M$, where $k_M$ is the comoving Hubble scale 
at which the radiation transits from UV to IR regime, 
is enhanced by the factor $\Delta = k/k_M$,
compared to the case of ordinary dispersion relation.
The transition wavenumber $k_M$ is estimated as
\beq
	k_M \simeq 2.6\times 10^3~{\rm Hz} \left( \frac{g_{*s}(T=M)}{106.75} \right)^{1/6}
	\left( \frac{M}{10^{11}~{\rm GeV}} \right),   \label{kM}
\eeq
where $g_{*s}$ denotes the relativistic effective degrees of freedom.
As we will see in the next section, 
$M$ is observationally bounded below as $M>10^{11}~$GeV
and hence the frequency (\ref{kM}) lies outside
the range covered by future space-based gravitational wave detectors,
such as BBO and DECIGO~\cite{Seto:2001qf}.
However, it may be possible to relax this lower bound as $M>10^7~$GeV
as explained in the next section.
In this case characteristic spectral shape may be observed by BBO/DECIGO.

\begin{figure}[t]
\includegraphics[scale=0.8]{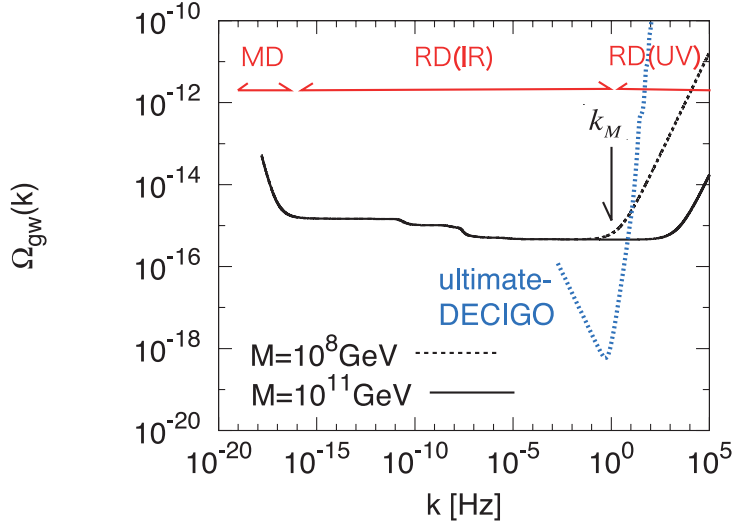}
\caption{Present gravitational wave background spectrum.
	Solid (dashed) black line corresponds $M=10^{11} (10^{8})$~GeV. 
	A mode enters the horizon in the matter dominated era, radiation dominated era
	with IR regime and radiation dominated era with UV regime as shown by
	MD, RD(IR) and RD(UV) on the top of the figure, respectively. 
	Sensitivity of ultimate-DECIGO is also shown by blue dotted line.}
\label{fig:GW}
\end{figure}

Fig.~\ref{fig:GW} shows the gravitational wave background spectrum.
Solid (dashed) black line corresponds $M=10^{11} (10^{8})$~GeV. 
A mode enters the horizon in the matter dominated era, radiation dominated era
with IR regime and radiation dominated era with UV regime as shown by
MD, RD(IR) and RD(UV) on the top of the figure, respectively. 
Sensitivity of ultimate-DECIGO is also shown by blue dotted line.
Here the tensor-to-scalar ratio is assumed to be 0.1.\footnote{
Throughout this section we have assumed that the gravity waves satisfy the ordinary relativistic
dispersion relation. This requires a scale $M_{\rm gw}$, above which the dispersion relation for
the gravity waves is modified like \REF{dis-rel}, to be larger than the Hubble parameter
during inflation. Since $M_{\rm gw}$ may be different from $M$, this is not a serious issue.
If $M_{\rm gw}$ is smaller than the Hubble parameter during inflation, the amplitude of the
gravity waves will be given by $\sim M_{\rm gw}/M_P$ instead of $H_{\rm inf}/M_P$,
according to the discussion along with Ref.\,\cite{Mukohyama:2009gg}.
Even if $M_{\rm gw}$ equals to $M$, the enhancement of the gravity waves still exists, although the
detection of the stochastic gravity waves by DECIGO becomes slightly difficult due to 
smaller tensor-to-scalar ratio.
}
It is seen that future gravitational wave detectors may have a chance to detect 
characteristic spectral shape of the gravitational waves if $M$ happens to be close to the
current observational bound.

We have considered two simple examples above, but the enhancement can occur in other situations.
For instance, if the dark matter (e.g. gravitino) is produced from the inflaton decay, its abundance
will be similarly enhanced.
Or, if preheating occurs soon after the inflaton begins to oscillate,
significant amount of gravitational waves may be produced
with peak frequency corresponding to the comoving Hubble scale after inflation
\cite{Khlebnikov:1997di,Easther:2006gt,GarciaBellido:2007dg,Dufaux:2007pt}.
These gravitational waves are also enhanced in a similar way.

\section{Observational constraint on the dispersion relation}
\label{sec:3}

In the previous section we have assumed that the radiation, including the standard-model particles,
are subject to the dispersion relation \REF{dis-rel2} at a temperature higher than $M$. There are
observational constraints on the possible deviation from the ordinary dispersion relation, using
the arrival timing of the high-energy gamma rays coming from far distant sources
\cite{AmelinoCamelia:1997gz},
such as gamma-ray bursts and active galactic nuclei.
In fact, the dispersion relation \REF{dis-rel} leads to the energy-dependent photon velocity,
\beq
	v = \frac{d\omega}{dk} = c^2 \frac{k}{\omega}\left( 
	 	1+2\kappa \frac{k^2}{M^2} +3\frac{k^4}{M^4} \right),
\eeq
where we have restored the speed of light $c$ in the IR limit.
In the IR limit ($k \ll M$), we obtain
\beq
	v \simeq c\left[ 1+\frac{3}{2}\kappa\frac{k^2}{M^2} +\left(\frac{5}{2}-\kappa^2\right)
	\frac{k^4}{M^4} \right].  \label{vphoton}
\eeq
If $\kappa$ is $O(1)$, the second term gives dominant contribution to the variation of the 
propagation speed.
According to the MAGIC collaboration~\cite{Albert:2007qk}, the lower bound on the scale $M$ 
in this case reads\footnote{More or less similar bound was also found by the H.E.S.S.~\cite{Bolmont:2009ej}
and Fermi~\cite{Abdo:2009zz} collaborations.}
\beq
M \;>\; 10^{11}\,{\rm GeV},
\eeq
for $\kappa = O(1)$. The essence of the Ho\v{r}ava-Lifshitz quantum gravity is the appearance of the
first term in the r.h.s. of \REF{dis-rel}, and so, the precise value of $\kappa$ 
may be different from order unity.
In particular, if $\kappa$ is negligibly small, 
it is $k^4$ term in Eq.~(\ref{vphoton}) that is subject to observational constraints.
In this case the constraint on $M$ becomes relaxed as\footnote{
	Even tighter constraint may come from observations of ultra high energy cosmic rays
	with energy higher than $10^{7}~$GeV, which may push
	the bound as $M>10^{11}~$GeV.}
\beq
M \;> \;10^7\,{\rm GeV}.
\eeq
Note that, since the coefficient of $k^4$-term in Eq.~(\ref{vphoton})
	  is positive, it is not possible to attribute the observed delay in the arrival time of gamma-rays to the
	  energy dependence in the speed of light.\footnote{
	  We thank Q.~G.~Huang for pointing out this issue.
}

In order to see the enhancement of the relic gravity waves by the future experiments,
$M$ must be smaller than $10^8$~GeV or so, as we have seen in the previous section.
This can be made compatible with the current observational
constraint by assuming $\kappa \ll 1$. On the other hand, the enhancement of the baryon asymmetry
can be viable for a broader parameter space.

\section{Conclusions}
\label{sec:4}

We have shown that the non-trivial dispersion relation suggested by the Ho\v{r}ava-Lifshitz
quantum gravity theory leads to a peculiar dependence of the radiation energy density on the scale
factor of the Universe. Such a dependence leads to an enhancement of the baryon asymmetry,
gravity waves, dark matter, and so on. We have demonstrated this enhancement
in simple examples. 

These phenomenological aspects are not limited to the Ho\v{r}ava-Lifshitz theory of gravity,
but applied to any quantum gravity model in which the dispersion relation
is modified at high energy.

\begin{acknowledgements}
Three of the authors (S.M., F.T. and S.Y.) thank K. Kohri for discussion.
The authors thank the Yukawa Institute for Theoretical Physics at Kyoto University, where this work 
was initiated during the YITP-W-09-01 on "Non-linear cosmological perturbations". 
This work is supported  by
World Premier International Research Center Initiative (WPI Initiative), MEXT, Japan.
S.M. is supported in part by MEXT through a Grant-in-Aid for Young
Scientists (B) No.~17740134, and by JSPS through a Grant-in-Aid for
Creative Scientific Research No.~19GS0219. 
K.N. would like to thank the Japan Society for the Promotion of
Science for financial support. 
S.Y. is supported in part by Grant-in-Aid for Scientific Research
on Priority Areas No. 467 ``Probing the Dark Energy through an
Extremely Wide and Deep Survey with Subaru Telescope''.
He also acknowledges the support from the Grand-in-Aid for the Global COE Program
``Quest for Fundamental Principles in the Universe: from Particles to the Solar
 System and the Cosmos '' from the
Ministry of Education, Culture, Sports, Science and Technology (MEXT) of
Japan.

\end{acknowledgements}



\end{document}